\documentclass[11pt,twoside]{article}

\usepackage{asp2014}

\aspSuppressVolSlug
\resetcounters

\begin{document}

\title{Reconstruction of IACT events using deep learning techniques with CTLearn}

\author{D. Nieto,$^1$ T. Miener,$^1$ A. Brill$^2$, J. L. Contreras,$^1$ T. B. Humensky,$^2$ and R. Mukherjee,$^3$ for the CTA Consortium}
\affil{$^1$Instituto de F\'{i}sica de Part\'{i}culas y del Cosmos and Departamento de EMFTEL, Universidad Complutense de Madrid, Spain; \email{d.nieto@ucm.es, tmiener@ucm.es}}
\affil{$^2$Physics Department, Columbia University, New York, USA; \email{aryeh.brill@columbia.edu}} 
\affil{$^3$Department of Physics and Astronomy, Barnard College, Columbia University, New York, USA} 

\paperauthor{D. Nieto}{d.nieto@ucm.es}{https://orcid.org/0000-0003-3343-0755}{Universidad Complutense de Madrid, Madrid, Spain}{Instituto de F\'{i}sica de Part\'{i}culas y del Cosmos and Departamento de EMFTEL}{Madrid}{Madrid}{E-28040}{Spain}
\paperauthor{T. Miener}{tmiener@ucm.es}{}{Universidad Complutense de Madrid, Madrid, Spain}{Instituto de F\'{i}sica de Part\'{i}culas y del Cosmos and Departamento de EMFTEL}{Madrid}{Madrid}{E-28040}{Spain}
\paperauthor{A. Brill}{aryeh.brill@columbia.edu}{}{Columbia University}{Physics Department}{New York}{NY}{10027}{USA}
\paperauthor{J. L. Contreras}{jlcontreras@fis.ucm.es}{https://orcid.org/0000-0001-7282-2394}{Universidad Complutense de Madrid, Madrid, Spain}{Instituto de F\'{i}sica de Part\'{i}culas y del Cosmos and Departamento de EMFTEL}{Madrid}{Madrid}{E-28040}{Spain}
\paperauthor{T. B. Humensky}{humensky@nevis.columbia.edu}{}{Columbia University}{Physics Department}{New York}{NY}{10027}{USA}
\paperauthor{R. Mujherjee}{rm34@columbia.edu}{}{Barnard College, Columbia University}{Department of Physics and Astronomy}{New York}{NY}{10027}{USA}



\begin{abstract}
Arrays of imaging atmospheric Cherenkov telescopes (IACT)
are superb instruments to probe the very-high-energy gamma-ray
sky. This type of telescope focuses the Cherenkov light emitted from
air showers, initiated by very-high-energy gamma rays and cosmic rays,
onto the camera plane. Then, a fast camera digitizes the longitudinal
development of the air shower, recording its spatial, temporal, and
calorimetric information. The properties of the primary
very-high-energy particle initiating the air shower can then be
inferred from those images: the primary particle can be classified as
a gamma ray or a cosmic ray and its energy and incoming direction can
be estimated.  This so-called full-event reconstruction, crucial to
the sensitivity of the array to gamma rays, can be assisted by machine
learning techniques. We present a deep-learning driven, full-event
reconstruction applied to simulated IACT events using
CTLearn. CTLearn is a Python package that includes modules for loading
and manipulating IACT data and for running deep learning models with
TensorFlow, using pixel-wise camera data as input.
\end{abstract}



\section{Introduction}

The ability of deep learning to assist in the analysis of data from imaging atmospheric Cherenkov telescopes (IACT) was first demonstrated by the detection of muon rings in real data~\citep{Feng_2016} and by the classification of gamma-ray and cosmic-ray simulated events ~\citep{Nieto:2017/K}. Subsequent studies proved its capability to reconstruct the energy and arrival direction of simulated gamma-ray events~\citep{2018arXiv181000592M, GammaLearnADASS20} and to improve IACT sensitivity on real data~\citep{2019APh...105...44S}. \texttt{CTLearn}\footnote{https://github.com/ctlearn-project/ctlearn}~\citep{Nieto:2019ak,ari_brill_2019_3345947} is a high-level, open-source \texttt{Python} package providing a backend for training deep learning models for IACT event reconstruction using \texttt{TensorFlow}. \texttt{CTLearn} allows its user to focus on developing and applying new models while making use of functionality specifically designed for IACT event reconstruction. The user can customize the training and built-in models hyperparameters. Hyperparameter optimization is available through the accompanying \texttt{CTLearn-optimizer} package. Data loading and pre-processing are performed using an associated external package, \texttt{DL1-Data-Handler}~\citep{bryan_kim_2020_3979698}. A diagram summarizing \texttt{CTLearn} architecture is shown in Fig.~\ref{fig:ctlearn}.

\articlefigure[width=.75\textwidth]{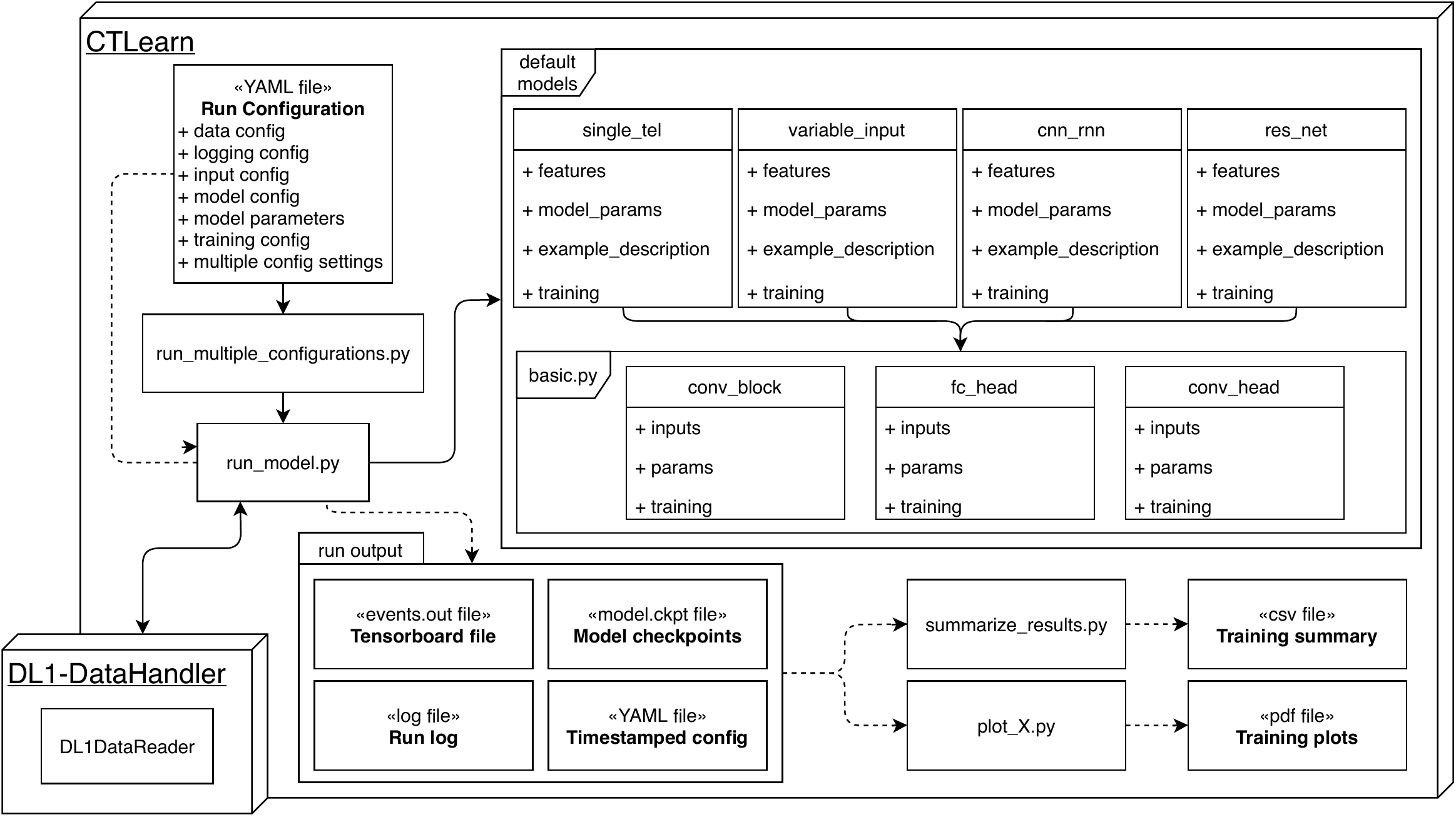}{fig:ctlearn}{Diagram summarizing \texttt{CTLearn}’s framework design.}

\section{Full-event reconstruction}

{\bf Model architecture.} \texttt{CTLearn} works with any \texttt{TensorFlow} model obeying a generic signature. In addition, \texttt{CTLearn} includes two built-in models for gamma/hadron classification of stereoscopic data~\citep{8909697} and a third one for full-event reconstruction of monoscopic data, dubbed \emph{TRN-single-tel} model (see Fig.~\ref{fig:trn-roc}, left panel). This last model is based on a deep convolutional neural network (CNN)-based architecture with residual connections \citep{2015arXiv151203385H} adapted from a thin ResNet \citep{2019arXiv190210107X}. A squeeze-and-excitation attention mechanism \citep{2017arXiv170901507H} was added into the CNN blocks. The architecture ends on a selectable fully-connected head that performs either particle classification or regression (energy or arrival direction reconstruction).

\articlefiguretwo{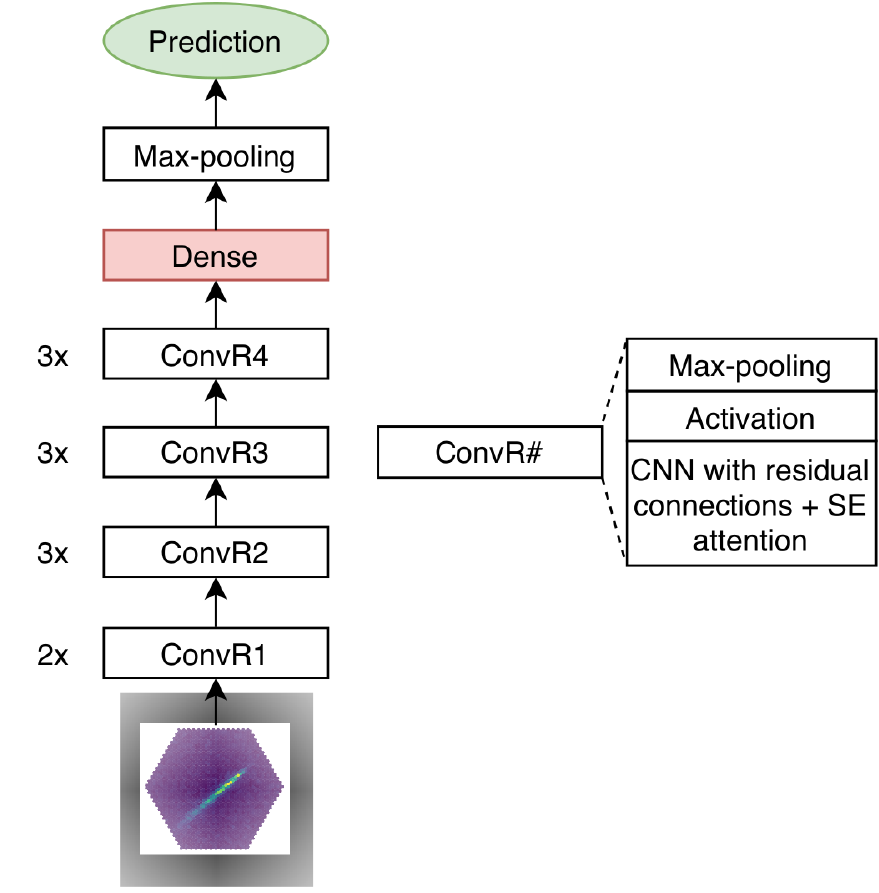}{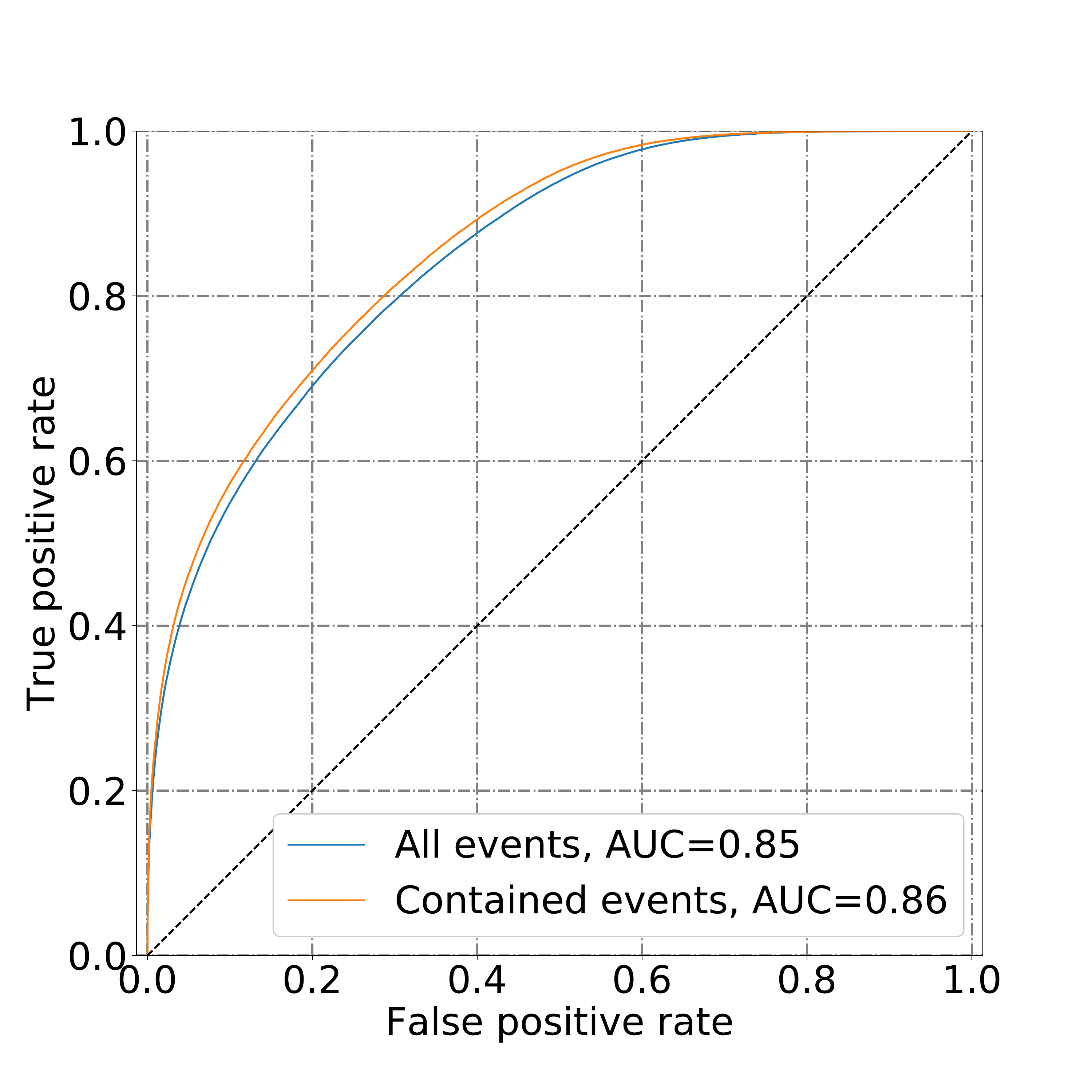}{fig:trn-roc}{\emph{Left)} \emph{TRN-single-tel} architecture. Prediction results are either particle type or energy or arrival direction. \emph{Right)} Example of ROC curves for the same given random seed.}

\noindent {\bf Experiments.} We trained and tested our \emph{TRN-single-tel} model with a dataset of simulated events for the Cherenkov Telescope Array\footnote{www.cta-observatory.org}~\citep[CTA]{CTAConsortium:2018tzg}, the next generation ground-based observatory for gamma-ray astronomy at very-high energies: ~ 2 million images from events having triggered an array of 4 large-sized telescopes (LSTs) in monoscopic mode
We considered diffuse gamma-ray and proton-initiated events, simulated within a cone of 10$^{\circ}$ radius - covering the whole field of view (FoV) of the instrument - in a balanced way. 
An 80\% of the data were used for training and a 20\% for testing. We trained the model on 200k batches of 64 images, validating periodically. The pixel layout in the original images is a hexagonal lattice, mapped to a Cartesian lattice using bilinear interpolation~\citep{Nieto:2019uj}. The model was trained independently for each reconstruction task. Two experiments were conducted: a) training and testing on all triggered events and b) training and testing on events where the images of the showers were mostly contained within the field of view of the telescope (no more than 20\% of the total image charge within the two outermost rings of pixels on the camera). Training was repeated 10 times for each experiment, setting a different random seed that varied the weight initialization and training set shuffling.\newline 
\noindent {\bf Results.} The \emph{TRN-single-tel} model successfully learned to perform full-event reconstruction across the entire FoV of the telescope. The classification accuracy (0.5 threshold) on the test set of diffuse events was $0.748\pm0.002~(0.756\pm0.001)$, with an area under the ROC curve of $0.848 \pm 0.002 ~(0.856 \pm 0.001)$ for the all-events (contained-events) experiment (average and standard deviation computed from the results of all training runs). An example of ROC curves can be found in Fig.~\ref{fig:trn-roc}, right panel. The energy resolution and bias, and angular resolution provided by the model on the test set of diffuse gamma-ray events can be found in Fig.~\ref{fig:results}, where the data points represent the median of all training runs and 16\% -- 84\% containment bands are shown, illustrating the inference robustness of the model. We remark that these results come from the reconstruction of diffuse events and note that the reconstruction for events showing arrival directions close to the center of the FoV would perform substantially better.

\articlefigurethree{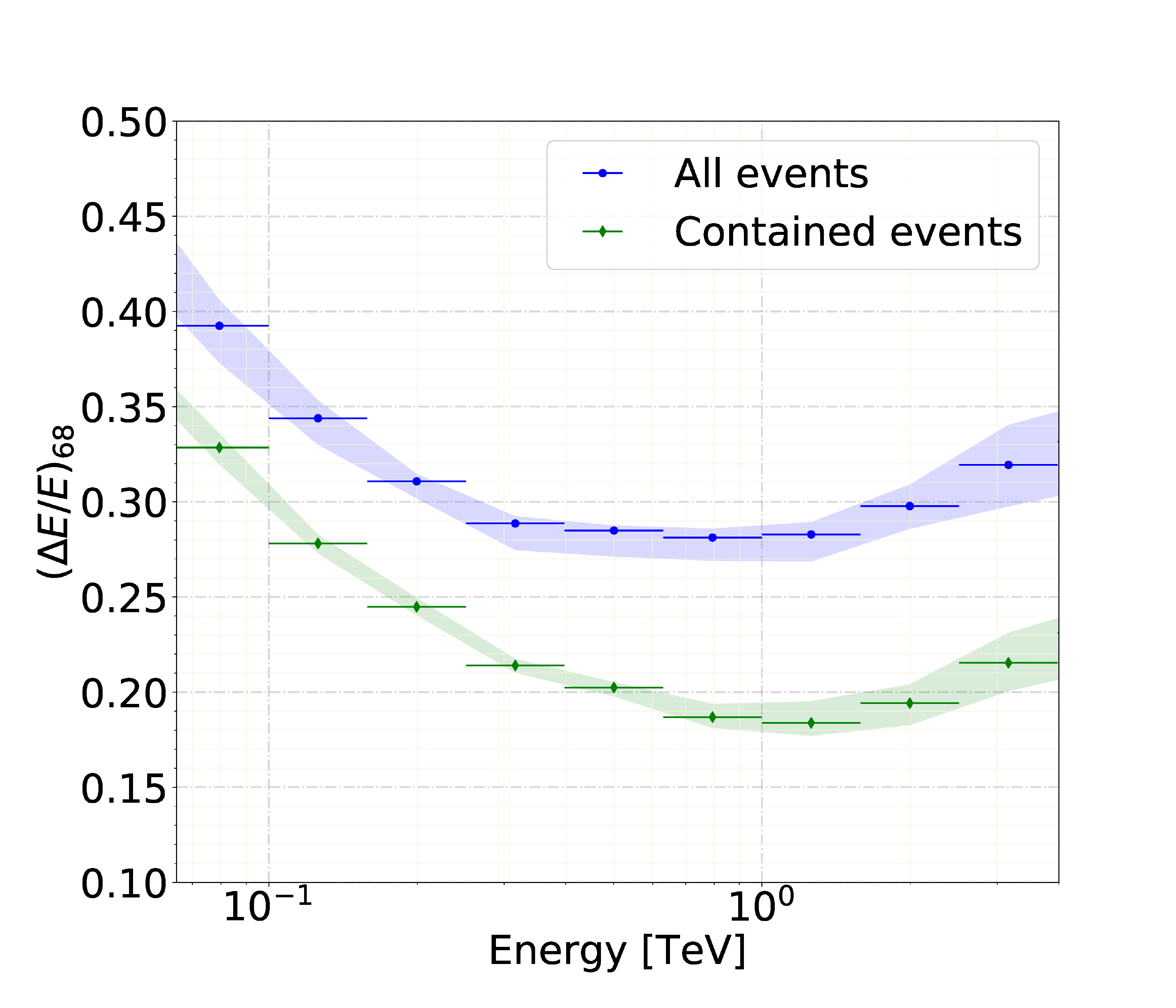}{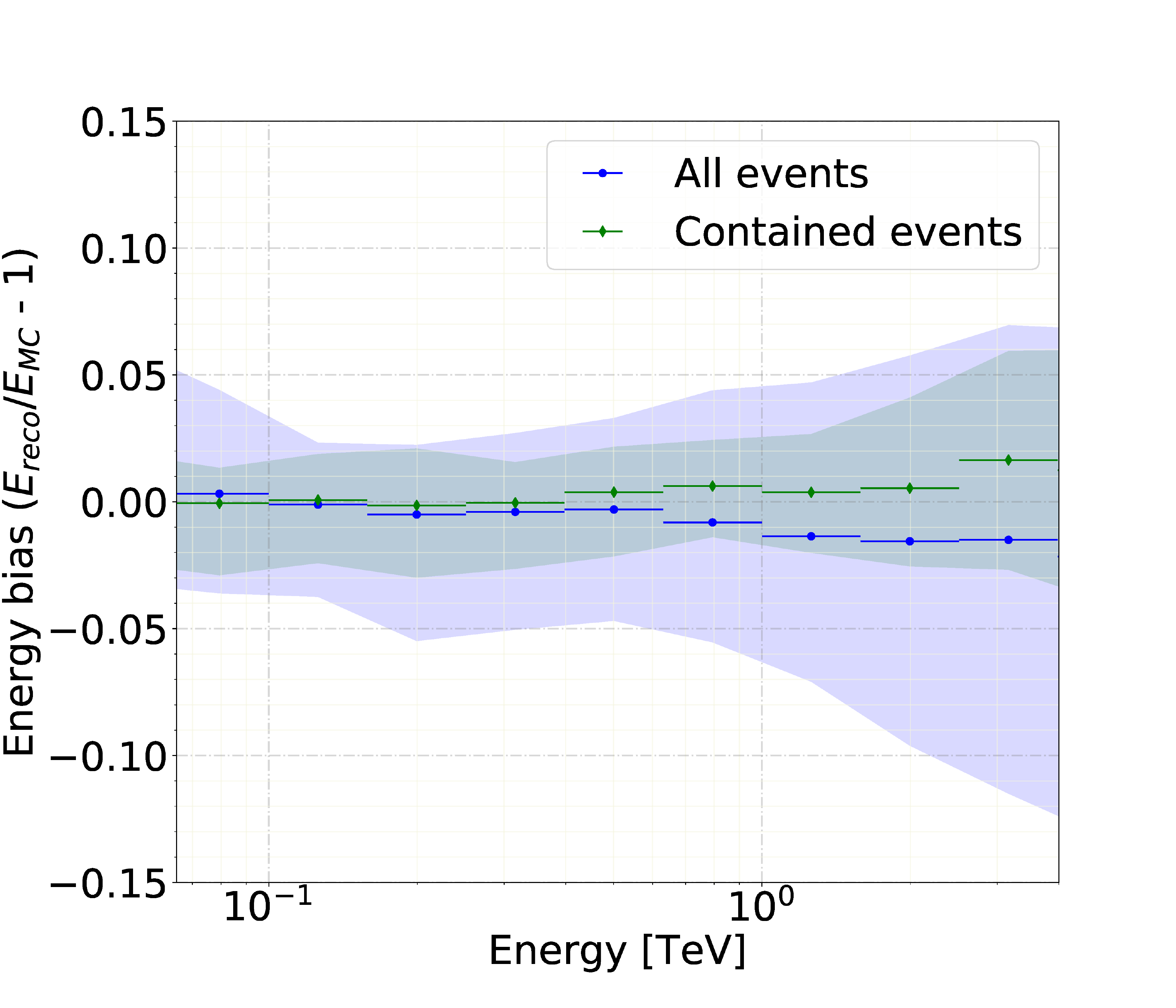}{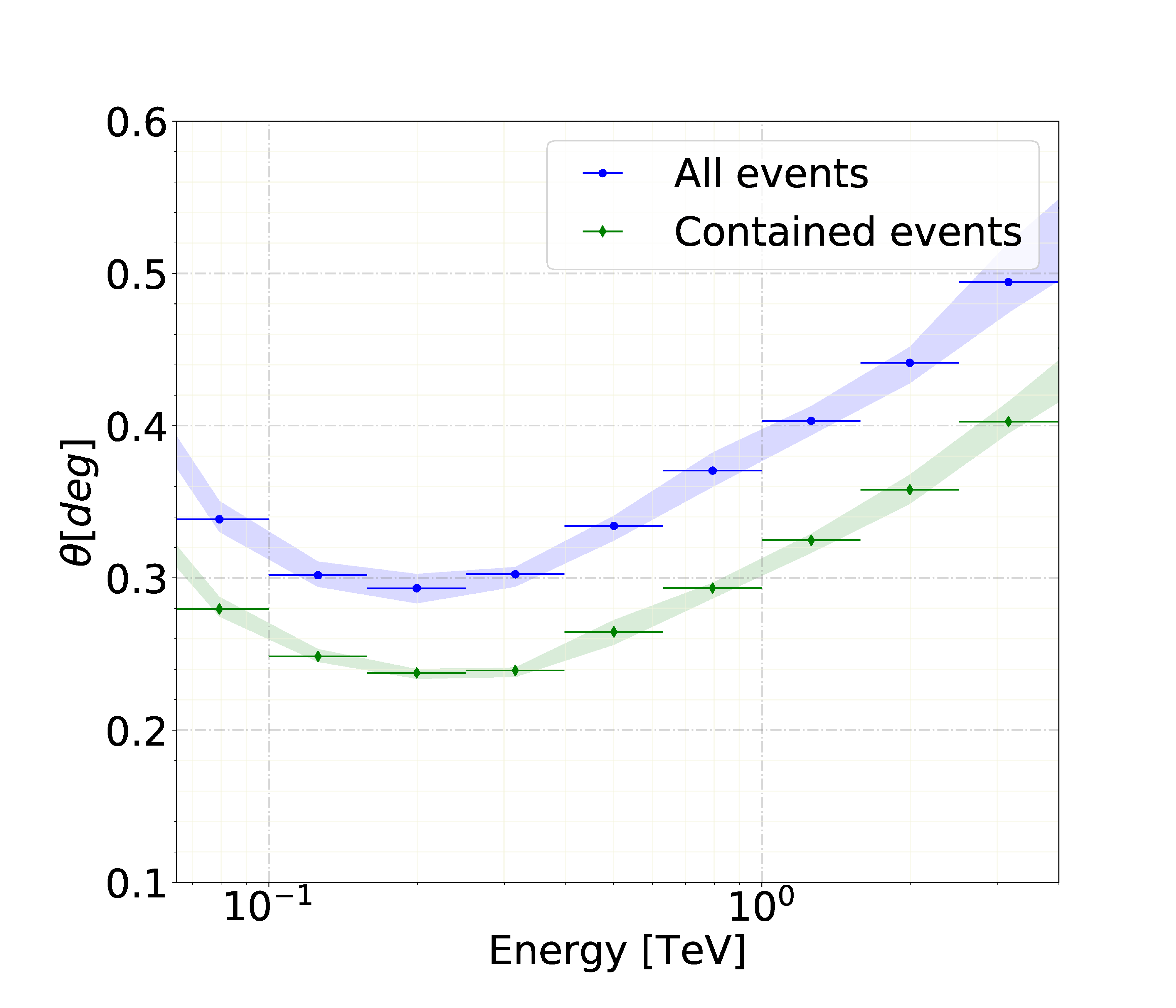}{fig:results}{\emph{Left)} Energy resolution; \emph{center)} Energy bias; \emph{right)} Angular resolution (all vs. reconstructed energy). Tested on diffuse gamma-ray events.}

\section{Conclusions and outlook}

\texttt{CTLearn}’s design and main features have been presented, showing its high level 
of configurability and flexibility, and demonstrating its potential for monoscopic 
full-event reconstruction using CTA simulated events. Areas where development is 
planned or already ongoing are: building models for stereoscopic full-event 
reconstruction; exploiting multitask learning; implementing models that could 
combine event-level data from a heterogeneous collection of telescope types, 
enabling IACT-specific metrics and loss functions; and ultimately applying 
full-event reconstruction to real IACT data.

\acknowledgements This work was conducted in the context of the CTA Analysis and 
Simulations Working Group. We gratefully acknowledge financial support from the 
agencies and organizations listed here:\\ http://www.cta-observatory.org/consortium\_acknowledgments.\\
DN and JLC acknowledge support from The European Science Cluster of Astronomy 
\& Particle Physics ESFRI Research Infrastructures  funded by the European Union’s 
Horizon 2020 research and innovation program under Grant Agreement no. 824064. 
TM acknowledges support from FPA2017-82729-C6-3-R. AB acknowledges support from 
NSF award PHY-1229205. DN acknowledges the support of NVIDIA Corporation with the 
donation of a Titan X Pascal GPU used for this research. The ASP would like to 
thank the dedicated researchers who are publishing with the ASP.

\bibliography{P5-48}  

\begin{thebibliography}{}
\expandafter\ifx\csname natexlab\endcsname\relax\def\natexlab#1{#1}\fi
\expandafter\ifx\csname url\endcsname\relax
  \def\url#1{\texttt{#1}}\fi
\expandafter\ifx\csname urlprefix\endcsname\relax\def\urlprefix{URL }\fi
\providecommand{\eprint}[2][]{\url{#2}}

\bibitem[{Acharya et~al.(2018)}]{CTAConsortium:2018tzg}
Acharya, B., et~al. (CTA Consortium) 2018, {Science with the Cherenkov
  Telescope Array} (WSP). \eprint{https://arxiv.org/abs/1709.07997}

\bibitem[{Brill et~al.(2019)}]{ari_brill_2019_3345947}
Brill, A., et~al. 2019, {CTLearn: Deep learning for imaging atmospheric
  Cherenkov telescopes event reconstruction}.
  \urlprefix\url{https://doi.org/10.5281/zenodo.3345947}

\bibitem[{{Brill} et~al.(2019)}]{8909697}
{Brill}, A., et~al. 2019, in 2019 New York Scientific Data Summit (NYSDS), 1.
  \urlprefix\url{https://doi.org/10.1109/NYSDS.2019.8909697}

\bibitem[{Feng \& Lin(2016)}]{Feng_2016}
Feng, Q., \& Lin, T. T.~Y. 2016, Proceedings of the International Astronomical
  Union, 12, 173–179.
  \urlprefix\url{http://dx.doi.org/10.1017/S1743921316012734}

\bibitem[{{He} et~al.(2015)}]{2015arXiv151203385H}
{He}, K., et~al. 2015, arXiv e-prints, arXiv:1512.03385.
  \eprint{https://arxiv.org/abs/1512.03385}

\bibitem[{{Hu} et~al.(2017)}]{2017arXiv170901507H}
{Hu}, J., et~al. 2017, arXiv e-prints, arXiv:1709.01507.
  \eprint{https://arxiv.org/abs/1709.01507}

\bibitem[{Jacquemont et~al.(2020)}]{GammaLearnADASS20}
Jacquemont, M., et~al. 2020, in these proceedings

\bibitem[{Kim et~al.(2020)}]{bryan_kim_2020_3979698}
Kim, B., et~al. 2020, {DL1-Data-Handler: DL1 HDF5 writer, reader, and processor
  for IACT data}. \urlprefix\url{https://doi.org/10.5281/zenodo.3979698}

\bibitem[{{Mangano} et~al.(2018)}]{2018arXiv181000592M}
{Mangano}, S., et~al. 2018, Lecture Notes in Computer Science,
  arXiv:1810.00592. \urlprefix\url{http://dx.doi.org/10.1007/978-3-319-99978-4}

\bibitem[{Nieto et~al.(2017)}]{Nieto:2017/K}
Nieto, D., et~al. 2017, in Proceedings of 35th International Cosmic Ray
  Conference {\textemdash} PoS(ICRC2017), vol. 301, 809.
  \urlprefix\url{https://doi.org/10.22323/1.301.0809}

\bibitem[{Nieto et~al.(2019{\natexlab{a}})}]{Nieto:2019ak}
--- 2019{\natexlab{a}}, in Proceedings of 36th International Cosmic Ray
  Conference {\textemdash} PoS(ICRC2019), vol. 358, 752.
  \eprint{https://arxiv.org/abs/1912.09877}

\bibitem[{Nieto et~al.(2019{\natexlab{b}})}]{Nieto:2019uj}
--- 2019{\natexlab{b}}, in Proceedings of 36th International Cosmic Ray
  Conference {\textemdash} PoS(ICRC2019), vol. 358, 753.
  \eprint{https://arxiv.org/abs/1912.09898}

\bibitem[{{Shilon} et~al.(2019)}]{2019APh...105...44S}
{Shilon}, I., et~al. 2019, Astroparticle Physics, 105, 44.
  \eprint{https://arxiv.org/abs/1803.10698}

\bibitem[{{Xie} et~al.(2019)}]{2019arXiv190210107X}
{Xie}, W., et~al. 2019, arXiv e-prints, arXiv:1902.10107.
  \eprint{https://arxiv.org/abs/1902.10107}

\end{thebibliography}


\end{document}